\documentstyle[pre,aps,epsf]{revtex}
\begin{document}
\bibliographystyle{prsty}
\twocolumn[\hsize\textwidth\columnwidth\hsize\csname@twocolumnfalse\endcsname
\title{Self-Organized Criticality in Deterministic Systems with Disorder}
\author{Paolo De Los Rios, Angelo Valleriani and Jos\'e Luis
Vega$^{\dagger}$}
\address{Max-Planck-Insitut f\"ur Physik Komplexer Systeme,
N\"othnitzer Str.\ 38, D-01187 Dresden.}
\maketitle

\begin{abstract}
\noindent
Using the Bak-Sneppen model of biological evolution as our paradigm,
we investigate in which cases noise can be substituted with a deterministic 
signal without 
destroying Self-Organized Criticality (SOC). If the deterministic
signal is chaotic the universality class is preserved; some 
non-universal features, such as the threshold, depend on the time correlation
of the signal. We also show that, if the signal introduced is periodic,
SOC is preserved but in a  different universality class, as long as the 
spectrum of frequencies is broad enough.
\end{abstract}
\vspace{1cm}
\vfill
]
\section{Introduction}
Due to Nature's inherent complexity, a lot of effort has gone
into developing mathematical models to describe it, 
if only qualitatively 
\cite{CJ89,Kauffman93,Ray94,Zaitsev92,PPVAC94,DMP91,CTP96,WW83,CR88,Sneppen92,Sneppen93,RRRIB93,MCFCB96,CGMRR96,MPB94,VDB96}. 
Among all natural processes, evolution has attracted a lot of attention
because of its global 
consequences  for life \cite{RN96,MP96}. 
Within the realm of evolution, one of the most fervidly argued topics is that 
of the explanation of mass extinctions \cite{Raup91}. 
Indeed, from a gradualistic point of view, mass
extinctions are rare events, 
due mainly to external abiotic factors 
such as earthquakes, meteorites
etc.\ \cite{Newman97}. From the point of 
view of {\em punctuated equilibrium}, on the other hand, 
mass extinctions are
bursts of activity between periods of stasis \cite{Raup86,GE77,EG88,EG72,BS93}.
The fossil record shows that the distribution of  mass extinctions follows
a power law \cite{SMBB97}. Among the many models proposed to 
describe evolution, those exhibiting Self-Organized Criticality (SOC)
\cite{BTW87,BTW88}  are of particular interest.
In layman terms, a system is called self-organized critical when 
it evolves towards a steady state in which certain 
physical quantities show fluctuations on any space and time scale (they follow
power law distribution).

In particular, we will concentrate on a model for macroevolution proposed by 
Bak and Sneppen (BS) \cite{BS93}. 
In it, extinctions are 
associated with avalanches of activity  without an inherent 
time or length scale.  In the original version of the model, no
influence of the environment was taken into account if not
implicitly in the fitnesses of the species. Later, Newman 
and coworkers \cite{Newman97} introduced a modified version of 
the BS model in which an environmental stress is introduced. 
All these versions of the model show self-organized criticality and
the randomness in the microscopic rule seems to play a relevant role.

In this paper we show that, if one substitutes  the random updating 
of the dynamic variable with a chaotic  or a periodic
map, SOC is not destroyed \cite{Note:4}. 
Some non-universal 
features will, however, depend on the time correlation of the signal. 
Moreover, if the signal introduced is periodic,
SOC is preserved but in a  different universality class, 
as long as the spectrum of frequencies is broad enough 
(a brief discussion
of some of these results can be also found in \cite{DVV97}).
Similar conclusions have been drawn in the context 
of standard fluctuation-dissipation processes by Bianucci 
{\em et.\ al.\ } \cite{Bianucci93,Bianucci94,BMFGW93}.
The paper is organized as follows. In Sec.\ \ref{BS} we review the main
features of the model as they are presented in the literature. After
introducing the maps in Sec.\ \ref{maps}, we show in Sec.\ \ref{models}
the results obtained for different deterministic 
updating rules. Our conclusions together
with a brief description of some open problems can be found
in Sec.\ \ref{concl}.

\section{General description of the model}
\label{BS}
The Bak-Sneppen model describes an  
ecosystem as  a collection of $N$ species on a $d$-dimensional
lattice,  each one of which 
can have $\cal{T}$ traits associated to it \cite{BP96}.
To each one of these traits corresponds a fitness described by a number $f$  
between $0$ and $1$ \cite{comment2}. 
Here, for simplicity, 
we consider the case with one trait, $d=1$, and periodic boundary conditions. 
To fix notation, we 
consider a one dimensional lattice of length $N$. The
 initial state of the system is defined by assigning to each site $j$ a 
random fitness $f^j_0$ chosen from a uniform distribution. 
The dynamics proceeds in three basic steps:
\begin{enumerate}
\item{} Find the site with the absolute minimum fitness 
on the lattice (this site
will be called the active site) and its two nearest neighbors.
\item{} Change, at the same time, the values of their fitnesses 
by assigning to  them new random numbers from a uniform distribution. 
\item{} Go to step 1.
\end{enumerate}
After an initial transient that  will be of no interest
to us, a non-trivial critical state is
reached. This critical
state, characterized by its statistical 
properties, can be understood as the {\em fluctuating balance} between two
competing ``forces''. Indeed, while the 
random assignation of the values, together with the coupling, 
acts as an entropic disorder, 
the choice of the minimum acts as an ordering
force. As a result of this competition, at the stationary state
the majority of the $f^j$ have values
above a certain threshold $f_c$. Only a few will be below $f_c$, namely those 
 belonging to the
running avalanche (see \cite{BS93,PMB96} for a detailed discussion).
Since the avalanches are the basic and fundamental mechanism of the model 
it is therefore worth describing them in more general terms.
Let us suppose that the system is already at stationarity
and let us find the minimum fitness, say $f_o < f_c$. We update it 
together with
 its two nearest neighbors (the actual value of the minimum does not
really matter). This updating creates disorder in a small region in
space, where most probably there are some lattice sites with $f^j<f_c$. 
Then, the new minimum will most probably be
among the last three sites changed. 
The active site most probably will be one of the two
nearest neighbors, thus affecting another site to the left or to the right.
In the following time-steps new sites will be touched by the avalanche of
mutations. Here one sees clearly 
the two aforementioned forces at work:
Disorder (every new value is chosen at random) and order (we decide to
mutate always the smallest). Since the equilibrium drives the threshold to the
value $f_c > f_o$, this means that the $f_o$--avalanche (an avalanche during
which all the selected minima are below $f_o$)
will eventually come to an end, 
in a finite time $s_{f_o}\equiv s_o$. During this process the
$f_o$--avalanche will also cover a certain number of lattice sites, 
{\em i.e.\ } it will also have a spatial size $n_{cov}^{f_o}\equiv n_o$. 
This feature 
gives the possibility to analyze and find the critical
values even without considering the system or lattice as a whole but simply
analyzing the statistics of $f_o$--avalanches \cite{Grassberger95}. 
Moreover, the avalanche dynamics shows  that, 
as long as $f_o$ is close to $f_c$ the average size
(both in time and space) of the related avalanches has to grow
considerably. Both averages will eventually become infinite as 
$f_o\rightarrow f_c$,
but this does not mean that all the avalanches are infinite (or of the maximal
space or time lengths allowed by the simulations).
These $f_o$--avalanches can be described by means of a 
distribution function \cite{BS93,PMB96}  

\begin{equation}
\label{a}
P_a (s,f_o) \, =\, s^{-\tau}\, F(s 
\,(f_c - f_o)^{1/\sigma})\, .
\end{equation}
In Eq.\ (\ref{a}) $s$ is the time-size of the avalanche and $F$ is some 
yet unknown scaling function that behaves  like
\begin{eqnarray}
\label{b}
\begin{displaystyle}
F(s\, (f_c-f_o)^{1/\sigma})\, 
\end{displaystyle}& = & \left\{
\begin{array}{ccc}
1 & \mbox{ as} & s\rightarrow 0  \\
0 & \mbox{ as} & s\! \gg\! p_o \equiv (f_c-f_o)^{-1/\sigma}
\end{array}
\right. \, .
\end{eqnarray}
The average duration of an $f_o$--avalanche is given by 
\begin{equation}
\label{c}
\bar{s}_o\, =\,  (f_c-f_o)^{-\gamma}\, ,
\end{equation}
where the exponent $\gamma$ is given in terms of the previously defined
exponents $\tau$ and $\sigma$ by
\begin{equation}
\label{d}
\gamma\, = \, \frac{2-\tau}{\sigma}\, .
\end{equation}
Numerical calculations provide 
good estimates for the value of the threshold as well as 
the two exponents $\tau$
and $\sigma$ \cite{BS93,PMB96} (see  Table 1). 
It is also useful to define other exponents that can be
easily obtained from numerical simulations. First  we
 consider the {\it first
return time} distribution $P_f(n)$, namely 
 the distribution of the times between two
consecutive updatings of the same site 
(when it is the minimum). Another
distribution function is the {\it all
 return time} distribution $P_{all}(n)$, namely the 
probability that a given site, active 
at time $t=0$, is active again at time $t$. In 
both cases, one defines the corresponding exponents by
\begin{eqnarray}
\label{h}
\begin{displaystyle}
P_f(n)\,\sim\, n^{-\tau_f} \end{displaystyle} 
& \,\,\,\,\mbox{ and }\,\,\,\, & \begin{displaystyle} 
P_{all}(n)\,\sim\, n^{-\tau_{all}} \end{displaystyle}\, .
\end{eqnarray}

In Table 1 we have  listed  the values of these exponents 
as they are given in the
literature. It is worth noticing that these exponents are not independent
quantities. Indeed, the scaling relations derived in \cite{BS93,PMB96} show 
that at
most two of them can be independent. However, using the master equation 
\begin{eqnarray}
\label{mas}
\begin{displaystyle}
(1-f_o)\frac{\partial P_a(s,f_o)}{\partial f_o}\end{displaystyle} &  = & 
 \begin{displaystyle} -P_a(s,f_o) n_o(s)\end{displaystyle} \\
& & \begin{displaystyle} +
\sum_{s_1=1}^{s-1} P_a(s_1,f_o) n_o(s_1) P_a(s-s_1,f_o)
\nonumber
\end{displaystyle} 
\end{eqnarray}
derived in \cite{Maslov96} and  the fact that
 $n_o(s)\sim s^{\tau_{all}}$ and 
$\sigma=1+\tau_a-\tau$ \cite{BS93,PMB96}, one 
proves that the only
independent exponent of the model is $\tau_{all}$ 
of Eq.\ (\ref{h}) \cite{MDM97}. From Eq.\
(\ref{mas}) one can also derive an infinite hierarchy of equations for the
moments of the distribution. The first equation in this hierarchy
\begin{equation}
\label{f}
\frac{\partial \log \bar{s}_o}{\partial f_o}\, = \,
 \frac{\bar{n}_o}{1-f_o}\, ,
\end{equation}
relates the exponents (\ref{d}) to the average number 
of covered sites; here  $\bar{n}_o$ is the non-universal
 average number of sites covered by the
$f_o$-avalanches. Putting Eq.\ (\ref{c}) 
into Eq.\ (\ref{f}) gives the so called 
$\gamma$-equation \cite{BS93,PMB96},
\begin{equation}
\label{g}
\gamma\, =\, \frac{\bar{n}_o}{1-f_o}\, (f_c-f_o)\, .
\end{equation}
For models belonging to the same universality class {\em i.e.} with the same
$\gamma$ this equation relates the non-universal quantities $\bar{n}_o$
and the threshold $f_c$. In particular, as we shall see in Sec.\
\ref{Bernoulli}, to a bigger $f_c$ must correspond a smaller $\bar{n}_o$ for
fixed $f_o$.

An interesting consequence of Eq.\ (\ref{g}) is that it is possible to change
$f_c$ while remaining in the same universality class. 
This can be
obtained by modifying the entropic tendency. Indeed substituting the random 
updating with a correlated chaotic system one introduces a correlation that
leads to an increase towards $1$ of the threshold. 
On the other hand, a greater correlation in the updating map means that the 
system spends more time in the same site, thus covering less sites in the 
same number of time-steps in comparison with a less correlated map. 
{From}  Eq.\ (\ref{g}) it is clear that as $\bar{n}_o$ decreases,
$f_c$ increases.\\
\begin{center}
\begin{tabular}{||c|c|c||} \hline
{\it quantity}\rule[-2mm]{0mm}{7mm} & \multicolumn{1}{c|}{\it value}  &
\multicolumn{1}{c||}{{\it error}}   \\ \hline\hline
\rule[-2mm]{0mm}{7mm}$f_c$ &$0.66702$ & $8$  \\ \hline
\rule[-2mm]{0mm}{7mm}$\tau$ &$1.073$ & $3$  \\ \hline
\rule[-2mm]{0mm}{7mm}$\sigma$ &$0.343$ & $4$  \\ \hline
\rule[-2mm]{0mm}{7mm}$\gamma$ &$2.70$ & $2$  \\ \hline
\rule[-2mm]{0mm}{7mm}$\tau_f$ &$1.58$ & $2$  \\ \hline
\rule[-2mm]{0mm}{7mm}$\tau_{all}$ &$0.42$ & $2$  \\ \hline
\end{tabular}
\end{center}
\begin{center}
{\bf Table 1} The first four values exponents are quoted from 
\cite{Grassberger95} while the last two from \cite{BS93,PMB96}.
\end{center}
\section{Maps}
\label{maps}
As we have seen before, the source of mutations in the 
Bak-Sneppen model is the presence of random noise in the system.  
Since a chaotic map may exhibit statistical 
properties similar to those of random noise, 
a similar competition between order and disorder
could be established when one substitutes random updating with 
chaotic updating. 
To understand the similarities as well as the differences
between the two kinds of updating,
in this section we discuss some general properties of maps. 

A deterministic map is a rule in which the new value of the variable 
is given by
\begin{equation}
f^j_{n+1}=F(f^j_{n}) \label{eq:general}
\end{equation}
with $F$ a deterministic function and $j$ the lattice site.  
In what follows, we will only consider maps of the unit interval onto itself
(usually called {\em unimodular maps}). 
Disregarding  periodic trajectories,  one can define 
several statistical quantities that are generally used 
to describe the properties of a generic sequence $\{f_{n}\}$. 

The first quantity of interest to us is the {\em invariant measure}, 
$\mu(f)$.  
Formally, 
the invariant measure for a unimodular map is defined by
\begin{equation}
\label{invm}
\mu_{f_0} (f) 
= \lim_{N \rightarrow \infty}\frac{1}{N} \sum_{i=0}^{N} \delta[ f -
F^{i}(f_{0})]\, .
\end{equation}
If $\mu_{f_0}(f)$ does not depend on the initial value $f_{0}$, 
the map is called ergodic (and one refers to the measure as 
 $\mu (f)$). 
If a system is ergodic,  
 time averages are equivalent to phase space averages,
and then the time average of any function
$g(f)$ can be computed as a phase space average via
\begin{equation}
<g> = \lim_{N \rightarrow \infty} \sum_{i=0}^{N} g(f_{i}) = \int_{0}^{1} \mu
(f) g(f) df \, .
\end{equation}
To describe the behavior of individual trajectories one needs more detailed
information provided by the {\em Lyapunov exponent} $\Lambda$. 
The  Lyapunov exponent measures
the average rate of separation in $f$-space of two given
trajectories per unit of time. It can be computed as
\begin{equation}
\Lambda= \lim_{N \rightarrow \infty} \frac{1}{N}\sum_{i=0}^{N-1} \log |\frac{d
F}{df}(f_{i})| \, .
\end{equation}
If a map has a Lyapunov exponent $\Lambda>0$,  this means that 
two trajectories will diverge from each other exponentially. In this case the  
map is called {\em chaotic}. 
This property has a very important
consequence: A very small perturbation in the initial condition will produce
a completely different outcome. 
Moreover, successive outcomes of a chaotic  map will 
behave like a stochastic variable (statistically speaking).
Finally, we will make use of the {\em autocorrelation
function} $C(m)$, defined as
\begin{equation}
\label{autoc}
C(m)= \sum_{n=0}^{\infty} (f_{n+m}-\overline{f})(f_{n}-\overline{f})\, ,
\end{equation}
where 
\begin{equation}
\overline{f}= \lim_{N \rightarrow \infty} \frac{1}{N} \sum_{i=0}^{N} F^{i}
(f_{0})\, .
\end{equation}
The function  $C(m)$ is a measure of how the
deviations from the average at time $i$ are related to the deviations from the 
average $m$ steps apart \cite{Schuster89}.
In particular, chaotic maps are expected to show
exponentially decaying autocorrelation functions, {\em i.e.}
\begin{equation}
C(m) \approx e^{-m/\tau}
\end{equation}
where $\tau$ is the {\em correlation time}.

We will now proceed to summarise the properties of the different 
maps we will be using in the Bak-Sneppen model.
Before continuing, it is worth mentioning that in principle, the
 case of random noise can be considered 
as a particular case of (\ref{eq:general})
in which $F(f_{n}) \equiv \psi(n)$ with $\psi(n)$ a
random variable with a uniform probability density \cite{comment1}.
\subsection{Bernoulli maps}
Let us start by considering the Bernoulli map \cite{Schuster89}, namely 
\begin{equation}
\label{bernoulli}
f^j_{n+1}=G_{r}(f^j_{n})=[rf^j_{n}] \, ,
\end{equation}
where $[f]$ stands for the value of $f$ modulus $1$ and $r 
\in {\bf N}, r>1$ is a constant. 
It has been shown (see \cite{Schuster89} and 
References therein) that this map has a uniform invariant measure
\begin{equation}
\mu^{(BM)}(f)= 1\, ,
\end{equation}
where the function $\mu_{r}(f)$ has been defined in Eq.\ (\ref{invm}).
Moreover this map is chaotic and is characterized by a Lyapunov exponent 
given by 
\begin{equation}
\Lambda^{(BM)} = \log r \, .
\end{equation}
For this map one can easily compute the time autocorrelation function, 
namely
\begin{equation}
C^{(BM)}(m) = \frac{1}{12 r^{m}}=\frac{1}{12}e^{-m \ln r }\, ,
\end{equation}
where $C^{(BM)}(m)$ has been defined in Eq.\ (\ref{autoc}) and the
correlation time is given by
\begin{equation}
\tau = \frac{1}{\ln r}\, .
\end{equation}
One sees 
that the correlation time decreases as $r$ increases.
This means that given two maps with different values of $r$,
the one with the bigger value of $r$ will be closer to true
random noise and then will de-correlate faster. As we shall see in the
following section, this last property is of crucial importance in order to
understand the differences between BS models with different Bernoulli maps. 

\subsection{Logistic map}

Let us now consider the logistic map (sometimes called Feigenbaum
map), namely
\begin{equation}
\label{feig}
f^j_{n+1}= \lambda f^j_{n}(1-f^j_{n})\, .
\label{feigenbaum}
\end{equation}
The reasons for studying this map are manifold. On the one hand, 
this map has already been considered in the context of biological
evolution models and population dynamics \cite{May74,May76,MO76,Abramson97} 
and can thus provide a possible deterministic interpretation
of the evolution inside every ecological niche. Moreover, it has been shown
that it describes the 
behavior of a wide variety of systems in nature \cite{CE80}. On the other
hand, it has
a  regime in which it is chaotic as well as one in which it is not,  
depending on whether $\lambda$ is bigger or less than the critical value
$\lambda_\infty\sim 3.56994$ \cite{Schuster89} (for $\lambda>\lambda_{\infty}$
there are windows in which the map is periodic; in this paper we will take
$\lambda$ outside these windows).

If we consider the particular case $\lambda = 4$, 
the invariant density for this map is given by
\begin{equation}
\mu^{(LM)}(f) = \frac{1}{\pi \sqrt{f(1-f)}}\, ,
\end{equation}
the Lyapunov exponent is 
\begin{equation}
\Lambda^{(LM)} = \ln 2
\end{equation}
and the correlation function is given by
\begin{equation}
C^{(LM)}(m) = \delta_{m,0}\, .
\end{equation}
The fact that this map is chaotic does not mean that the trajectory cannot be
written explicitly. Indeed, it is easy to see that
\begin{equation}
f_{n}=\sin^{2}(2^{n}c)\, ,
\end{equation}
with the initial condition $f_{0}=\sin^{2}(c)$, is a 
trajectory of the logistic map in the case of $\lambda = 4$.
\subsection{Tent Map}
To better illustrate 
the effects of time correlations in the updating, we will also need
the so-called ``tent'' map (a ``linear version''
of the logistic map), defined as 
\begin{eqnarray}
\label{tnt}
 f^j_{n+1} = \left\{ \begin{array}{ll} 
            2f^j_{n}  & f^j_{n} < \frac{1}{2}\\ \\
             2(1-f^j_{n}) & f^j_{n} > \frac{1}{2}
                      \end{array}
        \right. \, .
\end{eqnarray}
This map is chaotic with Lyapunov exponent $\Lambda^{TM} = \ln 2$. 
Contrary to the case of the logistic map, 
the invariant measure for the Tent map is uniform {\it i.e.}
\begin{equation}
\mu^{(TM)}(f)= 1\, ,
\end{equation}
with an autocorrelation function given by 
\begin{equation}
C^{(TM)}(m) = \delta_{m,0}\, ,
\end{equation}
like in the case of purely random noise.
\begin{figure}
\vspace{0.5cm}
\centerline{\epsfysize=6cm\epsfbox{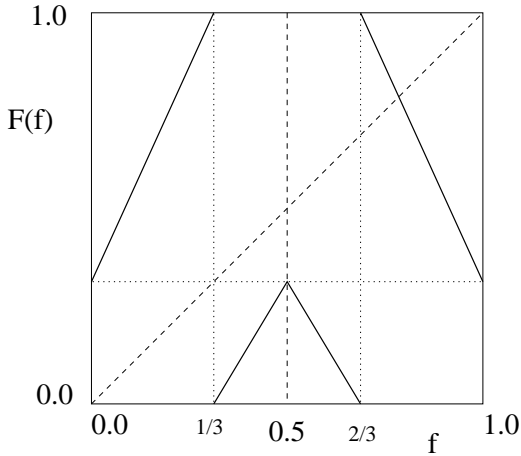}}
\caption{Shifted Tent map corresponding to $\eta=1/3$. 
In particular, for $\eta=0$ one
obtains usual Tent map, Eq.\ (\ref{tnt}).
\label{fig:tent_b}}
\vspace{0.5cm}
\end{figure}
For our applications (see next section), 
we find useful to define a modified version
of the Tent map (\ref{tnt}) (``shifted Tent map'')
in which we cut the $y-axis$ at $1-\eta$ and then shift the function $\eta$
up modulus one, as shown in the following equation

\begin{eqnarray}
\label{tntsh}
f^j_{n+1} & = & \left\{\begin{array}{cc}
                    2 f^j_n+\eta & 0<f^j_n<\frac{1-\eta}{2} \\ \\
                    2 f^j_n+\eta-1 & \frac{1-\eta}{2}<f^j_n<\frac{1}{2}\\ \\
		    \eta+1-2f^j_n & \frac{1}{2}<f^j_n<\frac{1+\eta}{2}\\ \\
		    \eta+2-2f^j_n & \frac{1+\eta}{2}<f^j_n< 1\, .
                    \end{array}\right.
\end{eqnarray}
As an example, in Fig. \ref{fig:tent_b} we can see 
the plot of Eq.\ (\ref{tntsh}) for the case $\eta = \frac{1}{3}$.
\subsection{Periodic Map}
So far we have only considered chaotic systems.
However, there are maps that are not chaotic but are ergodic. 
Let us consider a simple example of such a case in which the ``signal'' is
provided by an integrable system, that is a sequence given by 
\begin{equation}
f^j_{n} = \frac{\sin(\omega_j n + \phi_j )+1}{2}\, ,
\end{equation}
where the ${\omega_j},{\phi_j}$'s are the angular frequencies
 and  initial phases 
respectively. 
This can be rewritten as a map of $f^j_{n}$ onto $f^j_{n+1}$ as
\begin{equation}
f^j_{n+1}= \frac{\sin(\arcsin (2 f^j_{n}-1)+  \omega_j )+1}{2} \, ,
\label{oscillator}
\end{equation}
where the initial condition $f^j_0$ is given by 

\[f^j_0=\frac{1+\sin \phi_j}{2}\, .  \]

The invariant measure is not constant,  it is symmetric around $f=1/2$ and
peaked close to 
$f=0,1$
namely
\begin{equation}
\mu_{P}(f)=\frac{1}{2 \pi \sqrt{f^2-f}}\, .
\end{equation}
Since this ``signal''  is not chaotic, the Lyapunov exponent is zero and the
correlations will not decay exponentially. 
In fact, the correlations are given by
\begin{equation}
C^{(P)}(m) = \frac{\cos (\omega m )}{8}\, .
\end{equation}
At this point it is worth emphasising that these correlations are correlations
for a given sequence. If we consider two sequences with different 
values of $\omega, \phi$ the correlation will be different.
\section{Models}
\label{models}
Through the
use of the different maps presented in the previous section, we shall show here
that the random updating is no longer a necessary requirement to have
SOC. Moreover we will also show that as long as the map at hand is chaotic the
system does not change the universality class, {\em i.e.} all the exponents are
the same as in the case of random updating.

While the presence of
critical behavior was somehow expected, it is still surprising that
the universality class does not change. 
This means that the system is able
to self-organize at a higher level: It takes into account the temporal
correlation (or the average time spent in every site) by increasing
the threshold, so as to have the same statistical properties. What is even
more remarkable then is the fact that close enough to the threshold
it is not possible to distinguish the random updating case 
from the chaotic one from the microscopical point of view, 
being the statistical properties and all the variables exactly the same. 
As a consequence, all the equations and relations shown in Sec.\  \ref{BS}
are still valid for all the cases with chaotic updating.

We will show this equivalence through an  
infinite sequence of models with Bernoulli updating,
logistic and Tent map  
updating.
In fact, the same kind of analysis performed on the
case with the (modified) Tent map can show that the time
correlations are actually 
the ones responsible for the shifts in the thresholds.

However, the universality class is not always preserved. In fact, 
if one chooses
a non-chaotic updating rule the critical exponents may change. We will show
that by considering quasi-periodic updating rules (\ref{oscillator}).
\subsection{Bernoulli updating.}
\label{Bernoulli}
Let us consider a chaotic updating rule, whose statistical 
properties resemble those of a stochastic function, namely  the Bernoulli 
map, Eq.\ (\ref{bernoulli}).

\begin{figure}[h]
\centerline{\epsfysize=7cm\epsfbox{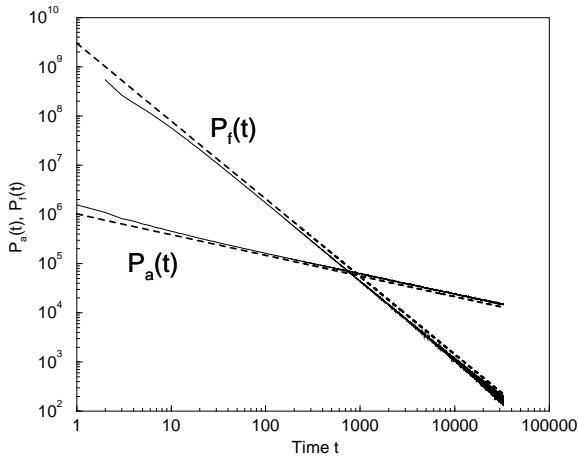}}
\caption{ First and all return distributions (non normalized) 
for a BS model with Bernoulli updating rule with $r=2$.  
For all the simulations shown here, we used a lattice of $2^{14}$  sites and 
$5\times 10^{9}$ iterations exploiting the tree-algorithm explained in
[35].
\label{fig:1a}}
\end{figure}

In Fig.\ \ref{fig:1a} we show the power-law behavior of the first and all 
return probability 
distributions in the case $r=2$.  The critical exponents  obtained coincide 
with those found in \cite{BS93,PMB96} for the random updating. 
Moreover,
for all values of $r$ the system falls in the BS universality class, 
{\em i.e.} it always has the same critical exponents. 
\begin{figure}[h]
\centerline{\epsfysize=7cm\epsfbox{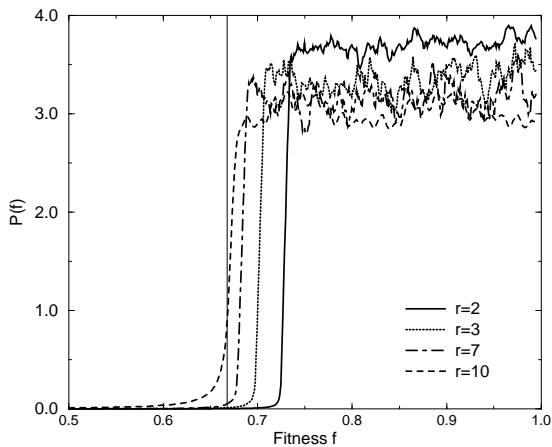}}
\caption{ Distribution of the fitnesses for $r=2,3,7,10$; the
threshold for $r=2$ is quite different from the usual BS threshold while 
the threshold corresponding 
to $r=10$ is very close to the BS value (given by the vertical line).
For all the simulations shown here, we used a lattice of $2^{14}$  sites and
$5\times 10^{9}$ 
iterations. 
\label{fig:1b}}
\end{figure}
The stationary distribution of the fitnesses, on the other hand,
follows a different pattern. Indeed, Fig.\ \ref{fig:1b} shows that the  
threshold for $r=2$ is bigger 
than the one found for the random case.
On increasing the value of $r$, the threshold moves towards the BS value (see
Fig. \ref{fig:1b}). 
For non integer values of $r$ ($r>1$), SOC is still preserved
within the BS universality class. However, in this case, 
 the distribution of the generated numbers is not
uniform and consequently it  
influences the distribution of the fitnesses at the stationary state.

\begin{figure}[h]
\centerline{\epsfysize=7cm\epsfbox{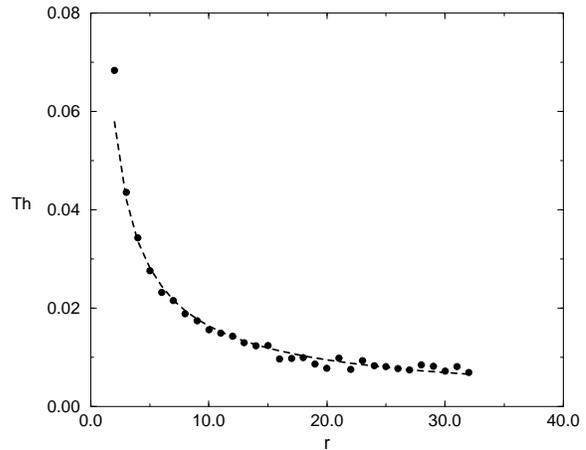}}
\caption{Values  of the thresholds $f_r$ for $r=2,\ldots,31$ after subtracting
the value $0.656$ for the random updating ($Th=f_r-f_{BS}$).
For all the simulations shown here, we used a lattice of $1000$  sites and
$5\times 10^{4}$ measurements.
\label{berna}}
\end{figure}

Turning now to Fig. \ref{berna}, we can see that the thresholds for the
Bernoulli updating approach the BS value as $r$ increases.
In Fig. \ref{berna} we have also
plotted the best fit we could find for the curve $(f_c(r)-f_c^{BS})$. This fit,
that corresponds to a power law $r^{-0.78}$, still remains an open
problem from the theoretical point of view.
There is a qualitative explanation for this behavior of the thresholds. As we
briefly mentioned in Sec.\ \ref{BS}, the change in threshold is an
indication of the correlation in time of the map we are using. Indeed, by
looking at Eq.\ (\ref{g}) we can see that an increase in
$f_c$ corresponds to a lower value of $\bar{n}_o$ for fixed $f_o$. This fact
means that the system spends more time per site and this reflects the fact that
it needs more time to de-correlate. At this stage one can also ask if 
Eq.\ (\ref{mas}) remains valid even with a correlated map and if it
is not necessary to introduce non-universal factors. 
The answer is given
by noticing that for fixed distance $\Delta_f$ from the threshold the
value $\bar{n}_{\Delta_f}$ is the same for all models, leading to
\begin{equation}
\bar{n}_{\Delta_f}\,=\, \frac{\gamma}{\Delta_f}\, ,
\end{equation}
where $\gamma$ is the universal exponent introduced in Eq.\ (\ref{g}) 
\cite{Maslov96}. 
This means that looking at
the system from a distance $\Delta_f$ from the threshold it is not possible to
distinguish two systems which have the same critical exponents.
\subsection{Logistic and Tent updating.}
In previous subsection we showed that models with time-correlated updating 
self-organize into a stable configuration with a threshold bigger than the
one in the random updating. At this stage, it is natural to consider
updating that, even if deterministic, are $\delta$-correlated. In particular,
we consider updating rules given by 
the logistic and Tent maps. 
\begin{figure}[h]
\centerline{\epsfysize=7cm\epsfbox{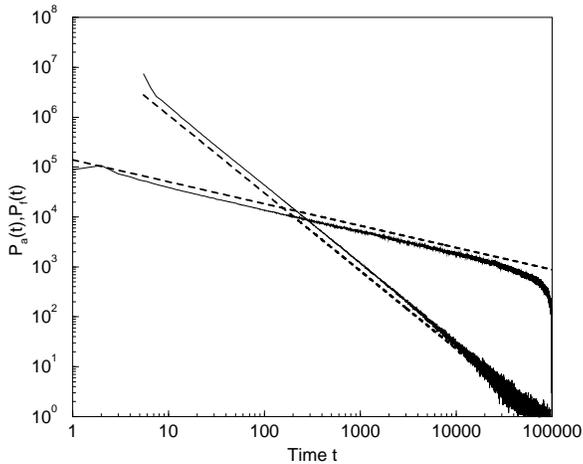}}
\caption{ First and all return distributions (non normalized) 
for a BS model with the
logistic map with $\lambda=4$ as updating rule. The exponents are the same as
for the BS model $\tau_f=1.58$ and $\tau_a=0.42$. In all the simulations shown
in this figure, we used a lattice 
of $2^{13}$ sites and $10^{8}$ iterations. 
\label{fig:2}}
\end{figure}
\begin{figure}[h]
\centerline{\epsfysize=7cm\epsfbox{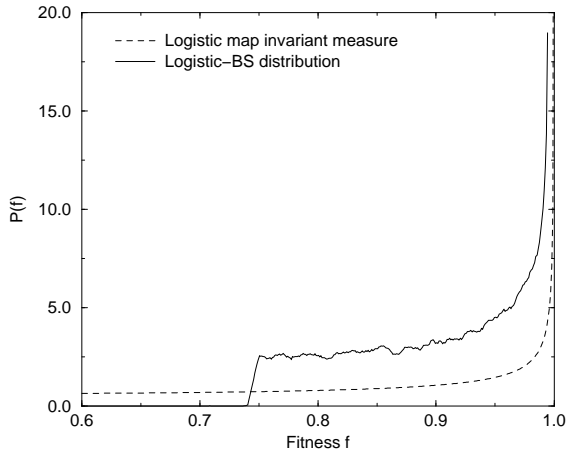}}
\caption{Distribution of the fitnesses for the logistic map. One can easily
see the effect of the non-uniform invariant measure.
\label{logfit}}
\end{figure}
Let us start by 
 taking  as updating rule for the
fitnesses the logistic 
map, Eq.\ (\ref{feigenbaum}).
As Fig. \ref{fig:2} shows, for  those values of $\lambda$ for which the map is
 chaotic,
 the system not only exhibits SOC but also stays in the same
universality class as the original BS model.
For $\lambda < \lambda_\infty$ we find that the system is not critical
any more. This is due to the fact that, for $\lambda < \lambda_\infty$ the 
map goes to a periodic orbit, and consequently the
updating is not ergodic. This case is 
equivalent to a BS model
with finite number of states for the fitnesses. In terms of our previous 
picture, the disorder force is too weak to ensure SOC.
\begin{figure}[h]
\centerline{\epsfysize=7cm\epsfbox{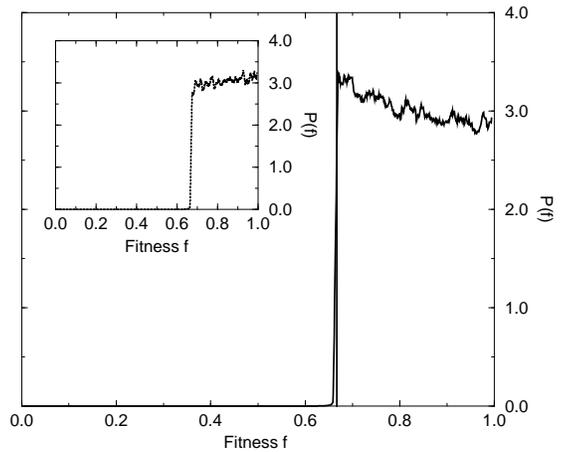}}
\caption{The main body of the figure shows the 
distribution of the fitnesses for the Tent map updating Eq.\ (\ref{tnt}).
The plot in the inset shows the fitness distribution for the case of the
shifted Tent map.
\label{fig:tent_updating}}
\end{figure}
One characteristic of the logistic updating is that, since the 
invariant measure is not uniform, the distribution of the fitnesses
above threshold is not uniform (see Fig.\ \ref{logfit}). 
This is not the case for
the Tent map, Eq.\ (\ref{tnt}) 
or the shifted version of it, Eq.\ (\ref{tntsh}).
For both these cases the fitness  distribution in the critical state is shown
in Fig.\  \ref{fig:tent_updating}. 
One observes that there is a peak in the fitness distribution
in the neighborhood of 
the threshold for Eq.\ (\ref{tnt}).
This can be understood as produced by the interplay between
the dynamics of the updating rule and the Bak-Sneppen dynamics.
Indeed, the Tent map has an unstable fixed point at 
$f^*=\frac{2}{3} \gtrsim f_{c}^{BS}$. Then if a site right below
threshold is chosen as the minimum, the updated value will be above threshold,
but still close to it. The next update in the same site 
will put the value of the fitness again below threshold, if a little bit
further apart \cite{Note:3}. Then, one needs to update this site several
times to remove it from the neighborhood of the threshold. 
Consequently, the probability of finding a site with fitness in the
neighborhood of $f_c$ is higher than in the random update.
If, on the other hand, one introduces the shifted Tent map, Eq.\
(\ref{tntsh}),  
where the fixed point is not close to $f_{c}^{BS}$, 
the distribution of the fitnesses above threshold 
is uniform, resembling exactly the random updating case 
(see Fig.\ \ref{fig:tent_updating}).

Comparing the different chaotic maps we can draw several conclusions. First,
time correlations in the updating immediately reflects in a shift of the
threshold in the sense that to higher correlations correspond higher
thresholds. Second, as shown by the shifted Tent map, the other
higher correlations do not in principle produce any measurable 
change in the statistical
properties of the system. 
\subsection{Periodic updating.}
Since time correlations in the updating rule do not, in principle, 
destroy SOC, it is worth considering systems in which the time
correlation of the updating does not
decay exponentially. As shown in \cite{DVV97}, the simplest example of
this class is given by a model in which the choice of the new fitness is
done according to the periodic map, Eq.\ (\ref{oscillator}).

As mentioned in Sec.\ \ref{maps}, choosing the initial 
phases is equivalent to
choosing the initial condition of the 
system. Consequently,  we take our phases $\phi_j$ 
at random ($0\leq \phi_j \leq \frac{\pi}{2}$). 
Our simulations indicate that if the frequencies are the same, that is
$\omega_j \equiv \omega$, the strong synchronization of the sites 
along the lattice destroys criticality (even though the fitnesses are
organized above a threshold). Indeed, the system develops 
a typical scale that is observed in the way of a cutoff 
in the distribution probabilities.
If we now choose  the frequencies $\omega_{j}$ such that
\begin{equation}
\omega_{j} \neq \omega_{i}
\label{disorder}
\end{equation} 
the situation changes dramatically. If we characterize 
the  frequency distribution by two numbers, namely
its centre $\omega_0$ and its width $\Delta\omega$,  
the behavior will indeed depend on both.
\begin{figure}[h]
\centerline{\epsfysize=7cm\epsfbox{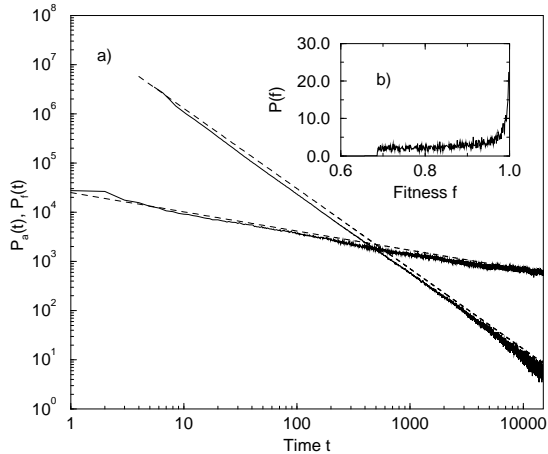}}
\caption{ (a) First and all return distributions (non normalized)
for a BS model with disordered
periodic updating rules; the exponents are $\tau_f=1.65(1)$ and
$\tau_a=0.38(1)$. (b) Distribution of the fitnesses. In all the simulations
shown in this picture, we used a lattice of $2^{12}$ sites 
and $5\times 10^{9}$ iterations. \label{fig:3}}
\end{figure} 
Even after long numerical investigation, the exact 
functional form of this dependence cannot be outlined 
in a satisfactory way. Nevertheless, it is clear that over a whole range of
values of the two parameters the system recovers a critical behavior. 
An
example can be seen in Fig. \ref{fig:3}, where we chose $\omega_0 = 19.5 \pi$
and $\Delta \omega = 19.5 \pi$.   
As mentioned above, the universality class 
changes with respect to the original BS model,
with $\tau_f = 1.65(1)$ and $\tau_a = 0.38(1)$, but the SOC behavior is
preserved. We observed that this universality class depends on the values of
$\omega_0$ and  $\Delta\omega$. For the sake of clarity, we show here only one
example, out of many, that illustrates the point. 
\section{Conclusions}
\label{concl}
Self-organized criticality in the BS model 
comes from the competition between 
the disorder in the updating  and the ordering pressure
of the minimum rule. For a given lattice and a given 
set of dynamical rules, the use 
of stochastic updating is tantamount 
to the introduction of  maximum disorder.
On the other hand, 
chaotic maps produce  series of numbers that 
resemble (statistically) pure random numbers, with 
the exception of the functional form of the invariant density and the
existence of decaying time correlations. The results presented here show that
the 
system, in its critical state, feels the details of the underlying 
dynamics, even if preserving the universality class.

The time correlations in the updating produce a change in the non-universal
features. In particular we showed that, as these correlations increase, the
critical state of the system  moves towards a more ordered configuration, that
is the threshold is higher. This correspondence is made evident, for example,
in the case of the shifted Tent map. There, a completely deterministic system
reproduces the original BS results. 

We would like to draw the attention of the reader to the complementarity
of the results presented here and those obtained by 
Bianucci {\it et. al.} 
\cite{Bianucci93,Bianucci94,BMFGW93}. They showed that if
 a variable $w$ (say a Brownian particle) is weakly coupled to a system, 
provided this system is chaotic or ergodic, the 
resulting deterministic motion of the variable $w$ conforms to
a standard fluctuation-dissipation process.
In fact, the irregularities of the deterministic 
statistics are washed out by the time scale separation between
the system of interest (represented by $w$) and the chaotic subsystem.
The chaotic system is referred to as a ``booster'' 
\cite{Bianucci93,Bianucci94,BMFGW93}.
This is completely analogous to what 
happens in the BS model. Noise (thermal or otherwise)
can be replaced by a deterministic system without significant
 changes in the stationary state. Stochasticity in the updating rule
is sufficient but not necessary:   SOC persists, even in the absence of chaos,
for (ergodic) periodic updating rules, if in a different universality class. 
Moreover, the conditions required from a deterministic system 
to be an appropriate booster are very similar to 
those required (from the updating rule) for  SOC to be preserved. 

Summarizing, the results presented
 here indicate that the feature ensuring SOC in systems with 
extremal dynamics, is not 
the randomness of the  updating  but the fact that the 
choice of the site  where the change is  performed 
(namely the minimum rule) is random. 
Moreover, as long as there is 
enough diversity among the species on the lattice,
the longer the memory (or the internal correlation) of
 each member, the higher the threshold. 
Indeed, in the case of chaotic maps, the diversity is ensured by the random
assignation of the initial 
values and as much as the chaoticity is increased we see that the threshold
decreases.
In the case of the periodic map instead, the random initial conditions do not
provide enough diversity. 
Indeed, in order to have SOC, the internal
time-scales, {\it i.e.\ } the periods, have to be distributed in a disordered
fashion. Briefly, one needs enough diversity for SOC to appear \cite{CPD97}.

At this stage,  several questions arise.
On the one hand, the behavior of the threshold with the parameter $r$ in 
the case of Bernoulli updating needs to be explained. Second, 
and perhaps more important from a theoretical point of view,
what is the exact relation between correlation time and the position of the
threshold?

Finally, we believe that these results add strength to the relevance of SOC in physics and biology,
since they allow different microscopic mechanisms to underlie
its appearance as a collective behavior.\\
{\em Acknowledgements:} We would like to thank R. Cafiero, P. Bak, S. Maslov,
and S.C. Manrubia for their useful comments and suggestions.


\begin{thebibliography}{99}
\bibitem{Note:4} In computer simulations one uses a deterministic map, 
   namely a pseudo-random number generator, to realize a
   random updating. Indeed,
   in any pseudo-random 
   number generator, a good deal of effort goes into designing the 
   algorithm to ensure that the statistical properties of the 
   outcome pass all the tests for randomness that any ``true'' random 
   sequence would. 
   So these  pseudo-random number generators are considered in practice, 
   even if strictly speaking deterministic, the best mimic for true 
   stochastic processes.
   For the purpose of this paper, the class of maps that will preserve SOC is
   much broader than the small set of good pseudo-random number generators 
\cite{Jansson66}.
\bibitem{comment2} Although this choice introduces some arbitrariness, we
believe that it is not possible to give an objective, generally valid
definition of fitness.
This means that sometimes the less fit species might be the one
with less offsprings while in some other circumstances the relevant parameter
could also be something else \cite{RJ86}. What the BS
model basically tells us is that at any given time the less fit species is
the one that is forced to change, whatever the meaning of fitness might be at
that time \cite{Lande85}.
\bibitem{comment1} There are other quantities (such as  the Kolmogorov-Sinai
entropy) that are also useful in studying general properties of maps but we
will not discuss here for lack of space.  The reader is kindly invited to look
at reference \cite{Schuster89} and references therein for a detailed
description of several of them.
\bibitem{Note:3} Note that points close to the fixed point of 
the map (\ref{tnt}), jump, upon iteration, 
from below to above $f^*$ and vice versa.
\end{thebibliography}
\end{document}